\documentclass[showpacs,aps,floatfix,prb,reprint,superscriptaddress]{revtex4-1}
\usepackage{graphicx}
\usepackage{xfrac}
\usepackage{amssymb}
\usepackage{tikz}
\usepackage{amsfonts} 
\usepackage{amsmath} 
\usepackage[utf8]{inputenc} 
\usepackage{mathtools}
\usepackage{xcolor,colortbl}
\usepackage{bm} 
\usepackage{array}
\usepackage{color}  
\newcommand{\f}[1]{\mathbf{#1}}

\newcommand{\br}{\mathbf{r}}

\newcommand{\bk}{\mathbf{k}}
\newcommand{\bR}{\mathbf{R}}
\newcommand{\bG}{\mathbf{G}}

\newcommand{\bra}[1]{\langle #1 |}
\newcommand{\ket}[1]{| #1 \rangle}
\newcommand{\braket}[2]{\langle #1 | #2 \rangle}
\newcommand{\psit}{\tilde{\psi}}
\newcommand{\phit}{\tilde{\phi}}

\begin{document}

\title{Accurate Tight-Binding Hamiltonian Matrices from \textit{Ab-Initio} Calculations: \\Minimal Basis Sets}
\author{Luis A. Agapito} 
\affiliation{Department of Mechanical Engineering and Materials Science, Duke University, Durham, NC 27708, USA}
\affiliation{Department of Physics, University of North Texas, Denton, TX 76203, USA}

\author{Sohrab Ismail-Beigi}
\affiliation{Department of Applied Physics and Center for Research on Interface Structures and Phenomena (CRISP), Yale University, New Haven, CT 06511, USA}

\author{Stefano Curtarolo}
\affiliation{Center for Materials Genomics, Duke University, Durham, NC 27708, USA}
\affiliation{Materials Science, Electrical Engineering, Physics and Chemistry, Duke University, Durham, NC 27708, USA}

\author{Marco Fornari}
\affiliation{Department of Physics, Central Michigan University, Mt. Pleasant, MI 48859}
\affiliation{Center for Materials Genomics, Duke University, Durham, NC 27708, USA}

\author{Marco \surname{Buongiorno Nardelli}} 
\email{mbn@unt.edu}
\affiliation{Department of Physics, University of North Texas, Denton, TX 76203, USA}
\affiliation{Center for Materials Genomics, Duke University, Durham, NC 27708, USA}
\date{\today}

\begin{abstract} 
Projection of Bloch states obtained from quantum-mechanical calculations onto atomic orbitals is the fastest scheme to construct \textit{ab-initio} tight-binding Hamiltonian matrices. However, the presence of spurious states and unphysical hybridizations of the tight-binding eigenstates has hindered the applicability of this construction.
Here we demonstrate that those spurious effects are due to the inclusion of Bloch states with low projectability. The mechanism for the formation of those effects is derived analytically. 
We present an improved scheme for the removal of the spurious states which results in an efficient scheme for the construction of highly accurate \textit{ab-initio} tight-binding Hamiltonians.
\end{abstract}
\maketitle 
\section{introduction} 

The tight-binding method, even in its simplest implementation, is a useful tool in the study of the electronic structure of molecules and solids.\cite{Huckel1931Model,Jones1934Tight_binding}
The advantage of the method is the tractable and intuitive understanding it affords by distilling the electronic structure  of complex systems into physically transparent Hamiltonian matrices expressed on a minimal basis set of atomic orbitals (AO).
For realistic materials, the tight-binding (TB) matrix elements have been typically calculated
by fitting to experiments or higher levels of theory. The resulting models were successfully applied in a panoply of complex materials with large supercells\cite{Papaconstantopoulos_JPCM_2003,Goringe_TightBinding_RPP1997,Colombo_Clusters_NATO1996}
  and for problems where a localized basis sets are  essential.\cite{Heine_ES_LAE_SS1980} A major shortcoming of the tight-binding approach is  the demanding fitting procedure that limits the application of the approach to well known materials and hinders the transferability 
 of the parameters to bonding environments outside the assumed training set.  

In recent years, the accuracy and reliability of the TB models have been largely improved with the introduction of \textit{ab-initio} tight-binding Hamiltonians. Here, the Hamiltonian resulting from a fully self-consistent quantum-mechanical calculation either within Density functional Theory (DFT) or other first principles approaches,  gets mapped into a much smaller space spanned by a set of atomic or atomic-like (i.e. Wannier functions) orbitals.

The representation of the electronic structure of the materials on a minimal TB basis set has been obtained with two main approaches:
(\textit{i}) The ``downfolding'' of the \textit{ab-initio} electronic structure (solved in the large basis) into a model containing only a few bands of interest which are disentangled from the rest;
 (\textit{ii}) the explicit calculation of  the matrix elements $\bar{H}_{\alpha \beta} =\bra{\phi_{\alpha}}\hat{H} \ket{\phi_{\beta}}$ using predetermined and fixed localized functions, typically AOs. 

In the first approach, one proceeds by selecting a subspace $\mathcal{B}$ (spanned by $N$ Bloch states $\ket{\psi_n}$ of interest)  of the $K$-dimensional space of the solutions of the original quantum mechanical problem. The latter is found by representing and diagonalizing the Hamiltonian $\hat H$ of the system using a very high-quality basis set of size $K$, \textit{e.g.} plane-waves with a large cut-off, a dense spatial grid, a large number of atomic-orbital-like Gaussian functions, etc. The subspace  $\mathcal{B}$ is then projected onto a space $\mathcal{A}$ generated by the atomic -like orbital functions, $\ket{\phi_\alpha}$ where $\alpha = 1, \dots, M$. Typically, the dimension of the $\mathcal{B}$ subspace, $N$ is much smaller of $K$, while the number of atomic orbitals is $M \geq N$ and is defined by the choice of the localized basis set in $\mathcal{A}$. Typically, in order to obtain a faithful representation of the electronic properties of the system in the smaller basis, the basis functions need to be iteratively optimized, thus adding a substantial computational effort.
Implementations of this approach include: muffin-tin orbitals of arbitrary order ($N$MTO) \cite{Andersen_muffintin_PRB2000}, maximally-localized Wannier functions (MLWF)\cite{Marzari2012MLWF}, quasi-minimal basis orbitals (QUAMBO)\cite{Lu2004QUAMBO}, etc.
While the optimized functions can be used to compute the TB matrix elements, their primary advantage is exploiting the information they contain to study the physics of the handpicked bands. For example, they can be used in mapping correlated bands into Hubbard models.\cite{Agapito2015ACBN0}  
Implementations of the second approach\cite{Horsfield1997Efficient_ab_initio_TB}  have used the non-self-consistent Harris-Foulkes\cite{Harris1985Simplified_method_for_the_energy,Foulkes1989Tight_binding_models} functional for $\hat{H}$ with the input charge density taken from the converged large-basis \textit{ab-initio} calculation.
The computational bottleneck in this approach is the calculation of multi-center integrals (three-center and up) for the functional. 
This approach can be readily extended to find the charge density self-consistently, thus, allowing efficient implementations of order-$N$ \textit{ab-initio} DFT codes.\cite{Sankey1989Multicenter_tight_binding_model,Soler2002SIESTA}

In this work we follow the principles outlined above but without resorting to an explicit basis set optimization. Using the  eigenstates $\ket{\psi_n}$ of $\hat{H}$ with $n=1,\ldots,N$, one can always write
\[
\bar H_{\alpha\beta} = \bra{\phi_\alpha}\hat H \ket{\phi_\beta} \approx 
\sum_{n=1}^N \braket{\phi_\alpha}{\psi_n}E_n\braket{\psi_n}{\phi_\beta}\, .
\]
Here the $\approx$ sign  is introduced because we restrict the sum to $N$ elements (the subspace $\mathcal{B}$) instead of $K$ (the ``complete'' basis within the limits of convergence). Defining the matrix of overlaps $B_{\alpha n} = \braket{\phi_\alpha}{\psi_n}$ and the diagonal matrix $E=\mbox{diag}(E_1,E_2,\ldots,E_N)$, the TB Hamiltonian matrix, $\bar{H}$, is expressed as
\begin{equation}\label{eq:tb_hamiltonian}
\bar H =BEB^\dag, 
\end{equation}
where $B$ is a  rectangular $M \times N$ matrix.
In this way, the computation of the TB matrix reduces to a straightforward matrix operation which  does not require any  special iterative procedure as needed by some of the methods discussed above in  (\textit{i}).
This construction takes advantage of the full knowledge of the eigenenergies and eigenfunctions obtained in the large-basis calculation, in contrast to using the charge density only as in (\textit{ii}). 
This scheme, also known as \textit{direct projection}, has been tried in the past but was considered unreliable:  even though it yielded an overall resemblance to the large-basis band structure, it introduces spurious states ``randomly'' scattered across the energy spectrum and unphysical hybridizations. 
See for instance the band structures in Fig.~\ref{fig:shifts4}b as well as Ref.~\onlinecite{Marzari2012MLWF} (Figs.~5 and 7, see also Ref.~\onlinecite{Teichler1971Best_localized_Wannier_functions})
\footnote{\label{note2} 
  Projection onto AO-like functions are used as the first iteration in the construction of MLWF functions. 
  TB band structures obtained with these trial Wannier functions are equivalent to those from the construction in 
  Eq.~\ref{eq:tb_hamiltonian}, see Ref.~\cite{Teichler1971Best_localized_Wannier_functions}} and  Figs.~1,2,4 in Ref.~\onlinecite{Agapito_2013_projectionsPRB}. 

We have previously shown that accurate TB Hamiltonians can be straightforwardly obtained from Eq.~\ref{eq:tb_hamiltonian} if only Bloch states $\ket{\psi_n}$ that project well  on the selected AO basis set (high projectability $> 95\%$) are included in the subspace $\mathcal{B}$, \textit{i.e.}, filtering. This process introduces a null space, which is shifted outside the energy window of interest.\cite{Agapito_2013_projectionsPRB} 
In this work we present a generalized scheme for the construction of TB Hamiltonians in a minimal basis set, suitable for cases when states with moderately high projectability ($ \gtrsim 85\%$) are needed to be included. The new scheme further enhances the accuracy of the TB Hamiltonian and has the added advantage of making the TB eigenvalues insensitive to the shifting operation. Furthermore, we use perturbation theory to analytically demonstrate that the spurious states and unphysical hybridizations previously observed in the {direct projection} scheme are due to the presence of low projectability states.

\section{Methodology}\label{sec:meth}

In our work we use  plane-waves (PW) as the large basis for the \textit{ab-initio} calculation of the Bloch states $\ket{\psi_n}$. The wave vector $\bk$ index is suppressed so the analysis for a periodic system can be understood to be at a particular $\bk$ point. 
We choose $\dim(\mathcal{A})=M$atomic-like localized orbitals $\ket{\phi_\alpha}$ (with $M \geq N=\dim(\mathcal{B})$)  which we assume to be an orthonormal set $\braket{\phi_\alpha}{\phi_\beta}=\delta_{\alpha\beta}$.  These could be Wannier functions or, more pragmatically, L\"owdin orbitals. The restriction to this subspace is obtained through the projector operator $\hat P = \sum_\alpha \ket{\phi_\alpha}\bra{\phi_\alpha}$ that is hermitian and idempotent. 

Let us consider a set of column vectors $\{\ket{B_n}\}$ obtained from direct projection of each Bloch wave $\ket{\psi_n}$ of the $\mathcal{B}$ subspace onto the chosen orthonormal atomic orbitals, $\hat{P} \ket{\psi_n}=\sum_{\alpha} B_{\alpha n} \ket{\phi_\alpha}$. 
The elements of the vector $\ket{B_n}$ are the projection coefficients $B_{\alpha n}=\braket{\phi_\alpha}{\psi_n}$. The explicit expression for the computation of these coefficients is given in Eq.~\ref{eq:proj4}. 

The projector $\hat{P}$ is represented as $P_B = B^\dag B$, a $N\times N$ matrix  with entries  
\[
(B^\dag B)_{nm} = \bra{\psi_n}\hat P\ket{\psi_m}.
 \]

The diagonal elements are the ``projectabilities'', defined as 
\[
p_n \equiv (B^\dag B)_{nn} = \braket{B_n}{B_n}=\bra{\psi_n}\hat P\ket{\psi_n},
\]
that measure to what extend the eigenstate $\psi_{n}$ is well-described in the space $\mathcal{A}$ specified by the projector $\hat{P}$.

While by construction $\mathrm{tr}(\hat{P})=M$, the trace of the matrix $P_B$ in $\mathcal{B}$ becomes 
\[
\mathrm{tr}(P_B)=\sum_{n=1}^N \bra{\psi_n} \hat P \ket{\psi_n} \leq N \leq M.
\]

If $\mathcal{A}$ is complete, all $\psi_{n}$ states project perfectly ($p_n = 1$) and the trace equals $N$, the size of $\mathcal{B}$. The deviation from $N$ is a criteria to assess the accuracy of the TB representation in $\mathcal{A}$ with respect to the electronic structure in the $\mathcal{B}$ subspace. 
\bigskip

High projectabilities are expected for the states in the lowest bands and poor projectabilities at higher energies. One would expect to obtain accurate TB eigenvalues and eigenvectors for the states with the largest $p_n  \approx 1$; however, as proved in Sec.~\ref{sec:tbh} states with low projectability, when folded in the TB Hamiltonian, hinder the accuracy of the results. 

In order to represent well the electronic structure of the system one needs to exclude from $\mathcal{B}$ the ``bad'' states with low projectability by choosing $N$ accordingly ($p_n$ larger than a chosen threshold for each $n=1,\dots, N$).
The procedure involves the construction  of the tight-binding Hamiltonian following Eq.~\ref{eq:tb_hamiltonian} with a normalized set of column vectors $\ket{A_n}=\ket{B_n}/\sqrt{p_n}$  such that $\braket{A_n}{A_n} = 1$. 

These states are used to build the initial TB Hamiltonian:

\begin{equation}
\bar H =A E A^\dag \, ,
\label{eq:HAH}
\end{equation}
where the $N$ columns of the matrix $A$ are the vectors $\ket{A_n}$. 
%
This product is an $M\times M$ matrix constructed using only $N$ states. Because of this construction, $\bar H$ is singular with an unphysical null space, $\mathcal{N}$, of size $M-N$ that compromises the accuracy of the eigenvalues (see Sec.~\ref{sec:tbh}).
%
%
In order to remove the effect of the null space we perform an orthogonal projection (see Ref. \onlinecite{Meyer2000LinearAlgebra}) using the set of vectors $\{\ket{A_n}\}$

\begin{equation}
{Q}_\mathcal{N}=I_M- A(A^\dag A)^{-1} A^\dag,
\label{eq:QA}
\end{equation}
where $I_M$ is the $M\times M$ identity matrix. 
%
Reconstructing 
the TB Hamiltonian as 
\begin{equation}
\bar H_\kappa = \bar H + \kappa Q_\mathcal{N},
\label{eq:Hk}
\end{equation}
is possible  to shift the eigenenergies corresponding to the null space elements to an arbitrary energy $\kappa$, away from the band with good projectability.
%
%
%
In practice, if only very high-projectability states are considered, $A^\dag A$ is close to the identity, and the shifting matrix can be approximated by $Q_\mathcal{N} \approx I_M -AA^\dag$, avoiding the matrix inversion. This approximation introduces a small $\kappa$ dependence into the states of the TB subspace. This dependence can be safely neglected when using a very high-projectability filtering criteria (\textit{e.g.} $p_n > 0.95$ in Ref.~\onlinecite{Agapito_2013_projectionsPRB}) or applying only small values of $\kappa$, otherwise the exact expression in Eq.~\ref{eq:QA} is required for
a faithful description of the energy bands.

\section{The Effect of low projectability states in the TB Hamiltonian}\label{sec:tbh}

Minimal basis set have proved satisfactory to achieve accurate TB matrices for periodic systems using the filtering procedure. 
However, if more unoccupied bands of high projectability are needed for a particular application, one can achieve that by progressively increasing the size of the AO basis set, \textit{e.g.} from single zeta (SZ) (minimal) to double zeta (DZ), etc. effectively increasing the size of $\bar{H}_\kappa$.
\footnote{L. A. Agapito, A. Ferretti, S. Curtarolo, M. Buongiorno Nardelli, \textit{in preparation}} 
Nonetheless, for most cases it is more advantageous to trade-off some accuracy away from the Fermi energy for the convenience of still dealing with TB matrices of smaller sizes, \textit{i.e.} to keep the basis minimal, especially in the study of systems with large number of atoms. This can be achieved by including bands with moderately high projectability ($p_n \gtrsim 0.85 $) in the construction of the TB matrix. However, as we  discussed in Sec.~\ref{sec:meth}, bands with low projectability affect the accuracy of the TB representation in $\mathcal{A}$.

To learn about the eigenvalues of $\bar{H}_\kappa$, we start by applying it to $A$.  One gets
\[
\bar{H}_\kappa A = \bar{H}A+\kappa Q_{\mathcal{N}} A = \bar{H}A= AEP\, ,
\]
where $P=A^\dag A$. To find an analytical expression for the eigenvalues of $\bar{H}_{\kappa}$, we assume the number of states $\psi_n$ to be equal to the number of AOs ($N=M$), so that $A$ is square and invertible. \footnote{\label{note1}  
This is possible as one can always compute more PW states to increase the number of columns of $A$. One can start with $M$ DFT states with the highest projectability $p_n$ to form $A$. If for some reason the corresponding $M$ columns vectors are not linearly independent, $A$ will not be invertible. In that case one can proceed in a organized iterative fashion: start with large pool of column vectors sorted by descending $p_n$, initialize a collection with the first vector and iteratively add another vector only if is linearly independent to the collection otherwise proceed to the next vector.} Then, the expression above,
$\bar{H}_\kappa A = A (EP)$, has a $M\times M$ square matrix $EP$ which is the representation of $\bar H$ in some linearly independent basis (columns of $A$) and its eigenvalues are also those of $\bar{H}_k$ (and $\bar H$) that we call $\bar E$.

$EP$, with the diagonal pulled out as a perturbation is then:
\begin{align*}
EP &= \mbox{diag}(P_{11} E_1,\ldots,P_{MM} E_M) + \\
&\left(
\begin{array}{cccc}
0         & P_{12}E_1 & P_{13}E_1    & \ldots\\
P_{21}E_2 & 0         & P_{23} E_2   & \ldots \\
P_{31}E_3 & P_{32}E_3 & 0            & \ldots\\
          & \vdots    &              & \vdots \\
          & \vdots    &              & P_{M-1,M}E_{M-1}\\
          &           & P_{M,M-1}E_M & 0 
\end{array}
\right).
\end{align*}

First of all, if any $p_n=1$, then that column and row of the perturbation are zero. The diagonal element is decoupled from the rest of the matrix so one eigenvalue will be exactly $\bar E_n = E_n$. In other words, a perfect representation of the exact wavefunction means the basis is complete for that state and that the action of $\hat H$ in that basis will be perfectly described.

Second, if one ignores the off diagonal entries, it can be seen that the eigenvalues of $\bar{H}_\kappa$ would be scaled versions of $\hat H$ where each eigenvalue is being scaled by its projectability so the energies of $\bar{H}_\kappa$ would be $\bar E_n \approx P_{nn}E_n$.  Therefore, bad projectability will incorrectly deliver a TB eigenvalue close to zero.  This has been the source of much trouble in previous TB methods without filtering.

Third, the off-diagonal elements of the perturbation matrix lead to hybridization. Namely, the lack of perfect projectability leads to level repulsion and further changes of the TB eigenvalues, beyond the scaling by $P_{nn}$ mentioned above. The perturbation matrix is small, since (\textit{i}) the off-diagonal elements of $P$ are negligible for all states with high projectability; and (\textit{ii}) they are still considerably small for the states with lower projectability (See Appendix~\ref{sec:offdiag}).
Second order perturbation theory on the off-diagonal elements gives the analytical expression for  the $n$-th eigenvalue of $\bar H$, which is called $\bar E_n$, changing from $P_{nn}E_n$ to

\begin{equation}
\bar E_n = P_{nn}E_n + \sum_{j \ne n}
\frac{ |P_{jn}|^2 E_nE_j}{P_{nn}E_n-P_{jj}E_j} + \mathcal{O}(P^3)\,.
\label{eq:pert_theory}
\end{equation}
In addition to showing that the hybridization changes the energies, the formula shows that the changes due to hybridization can be much larger than one would naively expect based only on looking at the small $P_{jn}$ entries: first the numerator has an additional $E_j$ energy factor, and second, the energy difference in the denominator is based on the scaled energies which means the energy difference can be smaller than between the actual eigenvalues (especially when both $P_{nn}$ and $P_{jj}$ are significantly smaller than one) thus enhancing the contribution of the hybridization.

\section{Results and discussion}
Here we illustrate the effects of low-projectability bands on the accuracy of the TB matrices derived for a molecular system, benzene, and a crystal, cobalt antimonide.
The \textit{ab-initio} calculations were performed with plane-wave DFT codes: \textsc{vasp}\cite{vasp} and/or \textsc{quantum espresso}.\cite{quantum_espresso_2009}

\subsection{Benzene}
In the case of benzene we computed the electronic structure using  \textsc{vasp} within the projector-augmented-wave (PAW) method\cite{PAW} and the Perdew-Burke-Ernzerhof (PBE)\cite{PBE} functional. We determined  the molecular wavefunctions $\psi_n$ (molecular orbitals) of an isolated benzene molecule in a large cubic supercell of 15 \AA\  side using  $\Gamma$-point to sample the reciprocal space and a kinetic-energy cutoff of 29.4 Ry.
 
The detailed procedure to compute the projection coefficients, $B_{\alpha n}$, is discussed in Appendix~\ref{sec:proj_bloch}. The states $\psi_n$  are projected onto a minimal basis set of $M=30$ AOs (C:$2s,2p$; and H:$1s$) taken from public repositories.\footnote{\label{note} The pseudo atomic orbitals used in this work are from public data sets:\\
\texttt{http://www.quantum-espresso.org/pseudopotentials/}\\
C.pbe-n-kjpaw\_psl.0.1.UPF \\
H.pbe-kjpaw\_psl.0.1.UPF \\
\texttt{http://qe-forge.org/gf/project/pslibrary/}\\
Co.pbe-n-kjpaw\_psl.1.0.0.UPF \\
Sb.pbe-n-kjpaw\_psl.1.0.0.UPF}

\begin{figure}[htb]
\begin{center}
    \includegraphics[width=0.99\columnwidth]{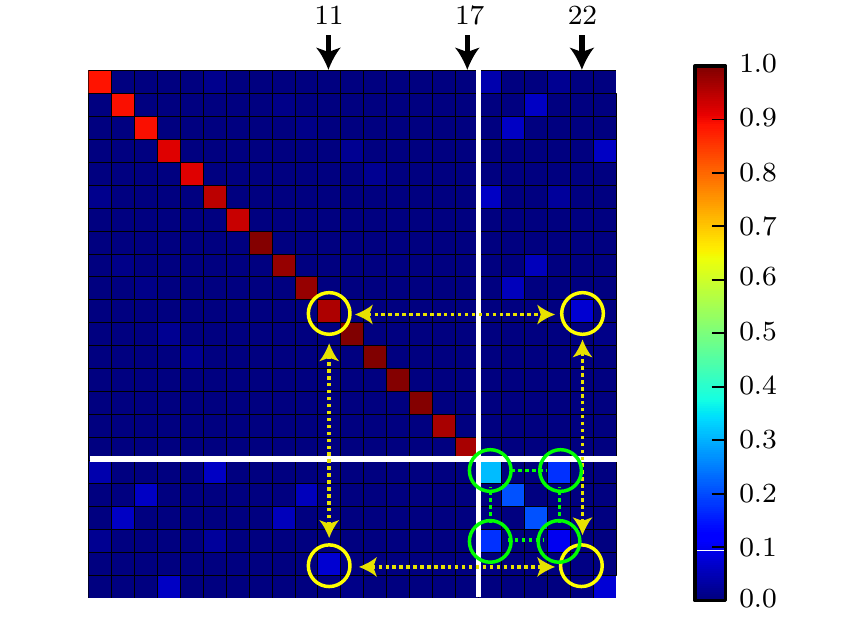}
    \vspace{-10mm}
    \caption{\small (Color online)  Projector matrix $|P_B|$ of benzene on a minimal AO basis set for the 23 molecular orbitals of lowest energy. The diagonal elements are the projectability numbers $p_n$. The presence of non-zero off-diagonal elements $\braket{B_m}{B_n}$ ($m \neq n$) reflects the non-orthonormality of the vectors $\ket{B_n}$.}
    \label{fig:Pmat} 
  \end{center}
\end{figure}

The diagonal elements of the projector matrix $P_B=B^\dag B$ shown Fig.~\ref{fig:Pmat} are the projectability numbers $p_n$ for each Bloch state $\psi_n$. In the chosen AO basis set, the 17 Bloch states of lowest energy have high projectability ($p_n > 0.88$) whereas states $18 \leq n \leq 20$ have low projectability ($0.20 < p_n < 0.32$). Moreover, higher-energy states ($21 \leq n \leq 23$) are not projectable in this particular AO set ($p_n=0.1,0.005,0.076$, respectively.)
\begin{figure}[htb]
\begin{center}
    \includegraphics[width=0.99\columnwidth]{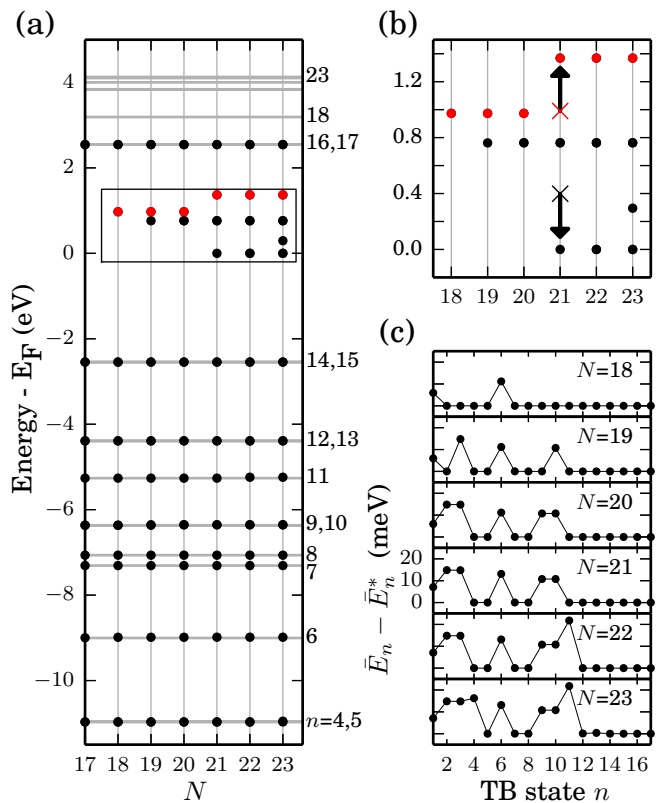}
    \vspace{-10mm}
    \caption {\small (Color online) \textbf{(a)} Evolution of the tight-binding eigenvalues with increasing number of low-projectability Bloch states. $N$ is the number of states used in the construction of the TB matrix, where the first 17 states are those of high projectability. In all case, the null states (not seen) are shifted by $\kappa = 8$ eV. \textbf{(b)} Zoom-in around the low-projectability TB eigenvalues. The red dots mark the TB eigenenergy corresponding to the 18th Bloch state .\textbf{(c)} Energy variation of each TB state $n$ against increasing $N$. The reference energy $\bar{E}^*_n$ corresponds to the Hamiltonian without low-projectability states ($N$=17).}
    \label{fig:shifts2} 
  \end{center}
\end{figure}
As discussed above, an accurate TB Hamiltonian matrix can be constructed by filtering out states with low projectability. Therefore, considering only the lowest $N = 17$ states ($p_n > 0.88$) yields TB eigenenergies that are in excellent agreement with the DFT values. The black dots in Fig.~\ref{fig:shifts2}a for $N$=17 show a maximum deviation from the DFT energies (gray lines) of only 5 meV.

\bigskip
In order to study the effect of states with lower projectability on the accuracy of the TB eigenenergies, we intentionally relax the filtering criterion to progressively include some states with lower projectability ($N>17$) in the $\mathcal{B}$ subspace. The vectors $\ket{A_n}$ corresponding to the low-projectability states are left unnormalized, \textit{i.e.},
\begin{equation}\label{eq:normal}
\ket{A_n}=\begin{cases}
\ket{B_n}/\sqrt{p_n},  & \mbox{if}\quad p_n \geq 0.85\\
\ket{B_n}            & \mbox{otherwise},
  \end{cases}
\end{equation}
The normalization of the high-projectability vectors $\ket{A_n}$ artificially makes the corresponding diagonal elements $(P_B)_{nn}=p_n\,(\approx 1)$ equal to 1, but this small change does not alter the analysis that follows. 

Expectedly, the TB eigenenergies $\bar{E}_n$ corresponding to low-projectability states largely underestimate the DFT values $E_n$. Those TB energies are seen in the zoom-in box in Figs.~\ref{fig:shifts2}a and b.   

The inclusion of the lower projectability state $n=18$, for instance, yields the ``bad'' TB eigenvalue $\bar{E}_{18}=0.9732$ eV. The large underestimation with respect to $E_{18}$, seen in Fig.~\ref{fig:shifts2}, is accounted by directly scaling the DFT values by the projectability, \textit{i.e.} $P_{18,18}E_{18}=0.9903$ eV, as discussed before. The evolution of $\bar{E}_{18}$ with increasing $N$ is shown in Fig.~\ref{fig:shifts2}b using red dots for visual aid. 

Including the degenerate states $n=$19,20 also yields strongly underestimated values $\bar{E}_{19}=\bar{E}_{20}=0.7621$ eV and in agreement with $P_{19,19}E_{19}=0.7877$ eV. 

Moreover, including the states $21 \leq n \leq 23$, which have even smaller projectabilities, yields TB eigenvalues close to zero: 0.0, 0.0, 0.2949 eV, respectively. Except for $\bar{E}_{21}$, these values compare well to $P_{nn} E_n =0.3984, 0.0218, 0.3117$ eV, respectively. The departure of $\bar{E}_{21}$ from $P_{21,21} E_{21}$ is due to hybridization effects discussed later in the text.

As hinted by Eq.~\ref{eq:pert_theory}, a state $\ket{A_n}$ hybridizes with another states $\ket{A_j}$ via a non-zero off-diagonal element of $P_{nj}=\braket{A_n}{A_j}$. 
The two diagonal blocks, of size $17 \times 17$ and $6 \times 6$, seen in Fig.~\ref{fig:Pmat}, correspond to high and low projectability states. The $17 \times 6$ and $6 \times 17$ off-diagonal blocks allow hybridizations between both types of states. Since all elements in the off-diagonal blocks are small ($\leq 0.0729$ eV), the low-projectability states are expected to have only a small impact on the high projectability ones. This is confirmed by inspecting the variations of the ``good'' TB eigenenergies ($\bar{E}_{n}$, $n=1\mbox{--}17$) while increasing the size $N$ of the subspace $\mathcal{B}$. The variations are small and not noticeable in Fig.~\ref{fig:shifts2}a. Instead, we plot the TB energies relative to reference values $\bar{E}^*_n$ in Fig.~\ref{fig:shifts2}c. $\bar{E}^*_n$ are the TB eigenenergies when $N$=17, that is, the high-projectability case. The maximum variation found is 21.8 meV and happens for $\bar{E}_{11}$ in the fifth panel (once state $n$=22 is included). This is consistent with the maximum element of the off-diagonal block happening at $|P_{11,22}|=0.0714$. We find that $P_{11,22}$ is the only non-zero off-diagonal element in the 11th row (and column) of the projector matrix, the energy variation can be directly attributed to the overlap between $\ket{A_{11}}$ and $\ket{A_{22}}$ following the hybridization mechanism depicted by the yellow circles and arrows in Fig.~\ref{fig:Pmat}. The variation is well estimated by the second-order perturbation model, Eq.~\ref{eq:pert_theory}, which reduces to:
\[
\Delta \bar{E}_{11} \approx  \frac{ |P_{11,22}|^2 E_{11} E_{22}}{P_{11,11}E_{11}- P_{22,22} E_{22}}= 21.7~\mbox{meV}\,.
\]

All hybridizations due to the off-diagonal elements in Fig.~\ref{fig:Pmat} (or similarly, of the matrix $P$) translate into peaks in Fig.~\ref{fig:shifts2}c. 
The number of peaks in each panel increases as more low-projectability states are progressively included. Every new peak $n$ that appears in a particular panel $N$ reflects the hybridization between a low-projectability state---the one newly introduced in panel $N$---and the ``good'' TB eigenstate $n$. Each new peak can be directly traced to a non-zero off-diagonal element in Fig.~\ref{fig:Pmat}. For instance, the peaks at $n$=1,6 in panel $N$=18 are due to $P_{1,18},P_{6,18}$; peaks $n$=3,10 in panel $N$=19 to $P_{3,19},P_{10,19}$; etc.
\bigskip

The elements of the off-diagonal blocks yield only small fluctuations; however, the overall maximum off-diagonal element $|P_{18,21}|=0.1757$ is inside the smaller $6 \times 6$ diagonal block. This indicates that hybridizations between the low-projectability states $\ket{A_{18}}$ and $\ket{A_{21}}$ are stronger.
Hybridization causes the level repulsion of $\bar{E}_{18}$ (red dot) along the upward arrow observed in Fig.~\ref{fig:shifts2}b at the introduction of $n$=21. The repulsion shifts up the level by 0.3953 eV from $P_{18,18}E_{18}$, which is the value expected in the absence of hybridization, marked by the red cross. The leading hybridization mechanism is depicted using green circles and lines in Fig.~\ref{fig:Pmat}. The perturbative estimate of the level repulsion is
\[
\Delta \bar{E}_{18} \approx  \frac{ |P_{18,21}|^2 E_{18} E_{21}}{P_{18,18}E_{18}- P_{21,21} E_{21}}= 0.6656~\mbox{eV}\,.
\]
$\bar{E}_{21}$ has the opposite level repulsion $\Delta \bar{E}_{21} = -\Delta \bar{E}_{18}$ (downward arrow), which explains the discrepancy between the value in the absence of hybridization $P_{21,21} E_{21}=0.3984$ eV (black cross) and the actual TB energy $\bar{E}_{21} \approx 0$ eV discussed earlier.

\begin{figure}[htb]
\begin{center}

    \includegraphics[width=0.99\columnwidth]{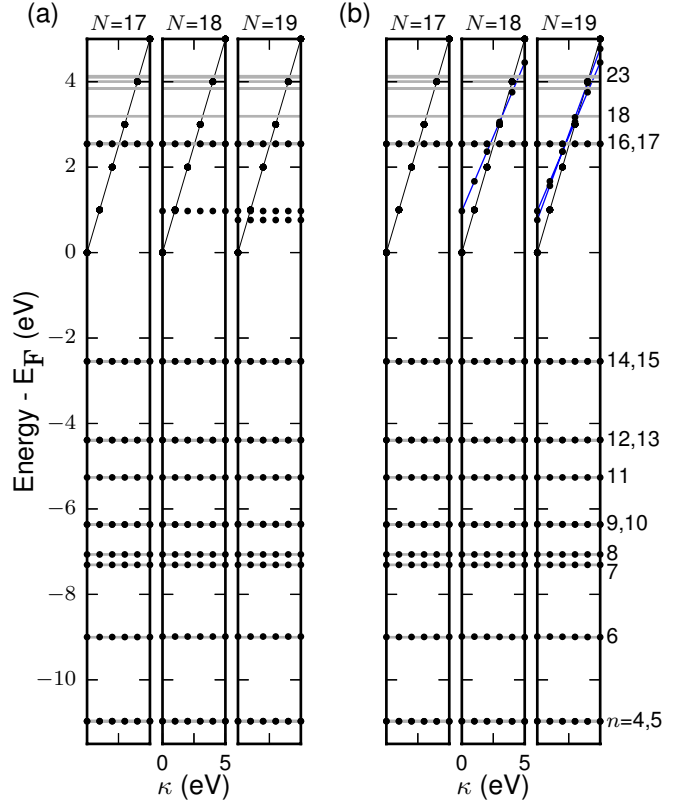}
    \vspace{-10mm}
    \caption{\small (Color online) Behavior of the null and low-projectability TB eigenstates under the shifting operation using \textbf{(a)} the exact $Q_\mathcal{N} = I_M - A(A^\dag A)^{-1} A^\dag$ or \textbf{(b)} the approximated $Q_\mathcal{N} \approx I_M - AA^\dag$ shifting matrix.}
    \label{fig:shifts} 
  \end{center}
\end{figure}


\bigskip

The construction in Eq.~\ref{eq:HAH} introduces a null space $\mathcal{N}$ containing $(M-N)$ degenerate eigenenergies $\bar{E}_\mathcal{N} = 0$ eV. The matrix $Q_\mathcal{N} = I_M - A(A^\dag A)^{-1} A^\dag$ is used to selectively move the null subspace upwards in energy by the control parameter $\kappa$, without affecting the remaining TB energies. With the exact $Q_\mathcal{N}$ the TB values do not acquire a dependence on $\kappa$. The evolution of the TB eigenvalues with $\kappa$ in Fig.~\ref{fig:shifts}a readily shows that only the degenerate null eigenenergies have a dependence on $\kappa$ (marked with solid black lines). The low-projectability TB states (dots about 0.85 eV in panels $N$=18 and $N$=19) do not belong to the null space and therefore are also independent of $\kappa$.
As argued in Sec.~\ref{sec:meth}, the shifting matrix can be approximated by $Q_\mathcal{N} \approx I_M - AA^\dag$, which avoids a matrix inversion, but at the cost of introducing a small $\kappa$ dependence to the good TB values. The approximation is safe when using a high-projectability filtering criterion ($N$=17). In this case both the exact (first panel in Fig.~\ref{fig:shifts}(a)) and the approximated $Q_\mathcal{N}$ (first panel in Fig.~\ref{fig:shifts}(b)) yield the same TB eigenvalues.
Nonetheless, the energy deviation due to the $\kappa$ dependence introduced by the approximated $Q_\mathcal{N}$ can become significant when low-projectability Bloch states are introduced, for instance, compare the second (and third) panels in Figs.~\ref{fig:shifts}(a) and \ref{fig:shifts}(b). With the exact $Q_\mathcal{N}$ (Fig.~\ref{fig:shifts}(a)) the low-projectability TB eigenvalues around 0.85 eV remain flat whereas they acquire a chiefly linear $\kappa$ dependence (marked with blue lines) when using the approximated $Q_\mathcal{N}$ (Fig.~\ref{fig:shifts}(b)). 



\begin{figure}[htb]
\begin{center}
    \includegraphics[width=0.99\columnwidth]{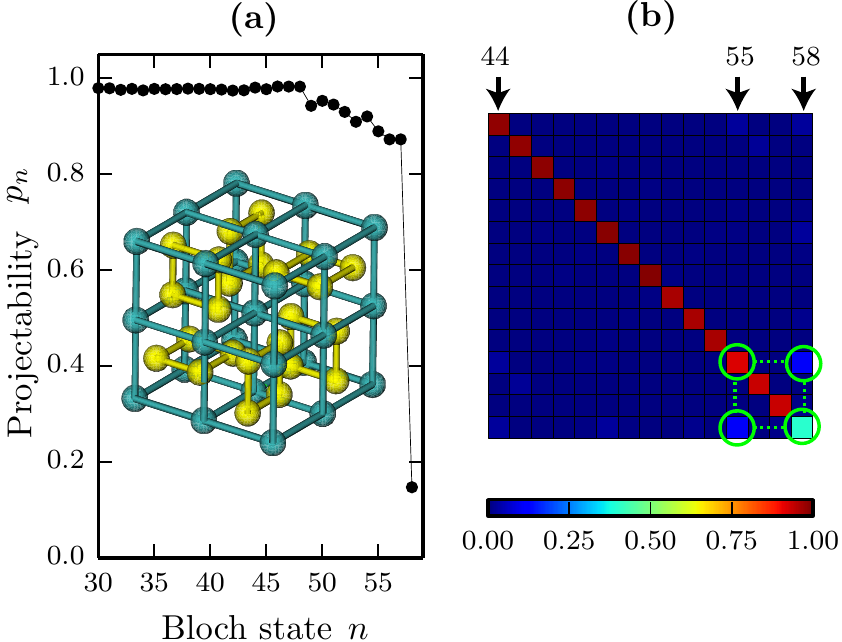}

    \caption{\small (Color online) \textbf{(a)} Minimum projectability $p_n$ per band $n$ over all $\mathbf{k}$ points for CoSb$_3$ on the chosen minimal basis set. A sharp decline of the projectability ($p_n < 0.2$) is seen for states $n \ge 58$. \textbf{(b)} Projector matrix $|P_B|$ at $\mathbf{k}_0=(0.4, -0.4, 0.5)$, in reciprocal coordinates. Only the matrix elements from $44 \leq n \leq 58$ are shown.}
    \label{fig:Pmat2} 
  \end{center}
\end{figure}

\begin{figure*}[th]
  \includegraphics[width=0.99\textwidth]{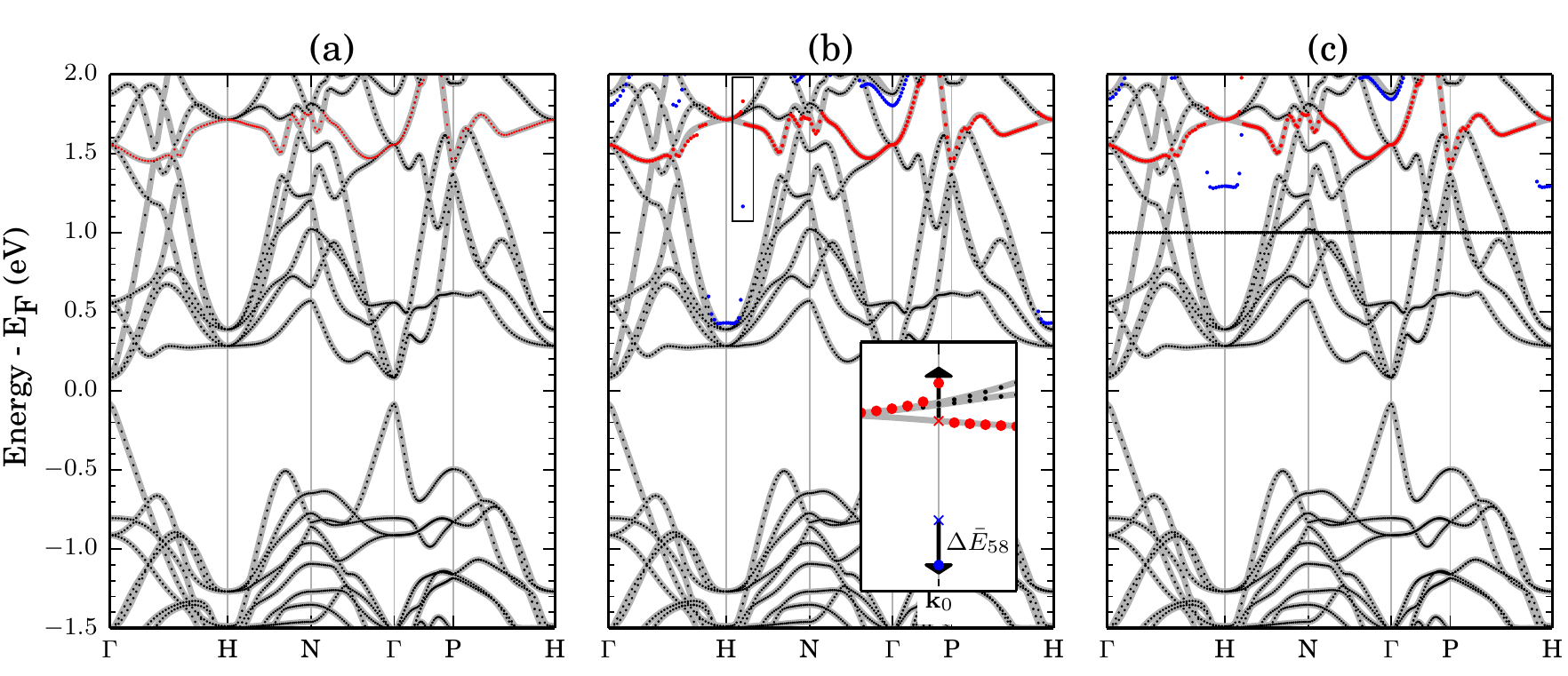}
  \caption{\small (Color online) Tight-binding eigenenergies for CoSb$_3$. \textbf{(a)} Accurate TB bands are obtained when adopting a high-projectability filtering criterion ($p_n > 0.87$) in the TB Hamiltonian. The band $n$=55 is shown in red. \textbf{(b)} The inclusion of a low-projectability band ($p_{58}=0.1474$) introduces the eigenenergies seen in blue, which induces unphysical hybridizations with the band $n$=55. The arrows in the inset, whose size is $|\Delta \bar{E}_{58}|$, illustrates the level repulsion at $\bk_0=(0.4,-0.4,0.5)$, in reciprocal coordinates, due to hybridization.  \textbf{(c)} The TB Hamiltonian is built with the approximated shifting matrix $Q_\mathcal{N}$ and $\kappa = 1.0$ eV to show the distinct dependences on $\kappa$ of the eigenstates. The reference DFT bands are shown with gray lines. Brillouin zone integration follows the AFLOW standard as discussed in Ref. [\onlinecite{Curtarolo:art58}]}
  \label{fig:shifts4}
\end{figure*}

\subsection{CoSb$_3$}
We analyze CoSb$_3$ as an example of a periodic solid. CoSb$_3$ is a typical binary skutterudite compound with cubic structure and space group \#204. Skutterudites are among the most promising thermoelectric materials.\cite{CoSb_natmat:2015}

We use the \textsc{quantum espresso} suite of \textit{ab-initio} codes to obtain the Bloch wavefunctions and the matrix $B$ of projection coefficients. The wavefunctions are obtained using the PBE functional with an energy cutoff of 50 Ry and the PAW data set from the PSlibrary 1.0.0. \footnotemark[2] We choose a minimal basis set to project onto composed of $M=84$ AOs (Co: $4s,4p,3d $ and Sb: $5s,5p$, taken from the PAW data set).

The projectability of all occupied DFT bands ($n \leq 48$) is very high $p_n > 0.97$ and progressively decreases for the unoccupied bands at higher energies (see Fig.~\ref{fig:Pmat2}(a)). The basis set supports 9 unoccupied bands with high projectability of $p_n > 0.87$ ($49 \leq  n \leq  57$) before declining to poor values of $p_n < 0.15$ for $n \geq 58$. 
An accurate TB Hamiltonian is obtained when considering only Bloch states of high-projectability ($N$=57) as confirmed by the excellent match between the TB (black and red dots) and the DFT bands (gray lines) seen in Fig.~\ref{fig:shifts4}(a). The band $n$=55 is shown in red.
\bigskip

To illustrate the effect of hybridization on the accuracy of the TB Hamiltonian, we include one Bloch state $n$=58 of low projectability ($p_{58}=0.15$) in the construction of the Hamiltonian. The corresponding TB band (not present in panel a) is shown in Fig.~\ref{fig:shifts4}b in blue.

First, it is seen that the TB band $n$=58 (blue dots in panel b) does not reproduce the reference DFT band, especially in regions of reciprocal space with the lowest projectability (around H) where the TB eigenvalues strongly underestimate the DFT values and are consistent with being scaled by their projectabilities, \textit{i.e.} $P_{nn} E_n \sim  0.4$ eV.

Second, the fidelity of the band $n$=55 (red dots) is noticeably reduced around H, with respect to panel (a), due to hybridization with the low-projectability band $n$=58 (blue dots), which leads to level repulsions.  
The inset in Fig.~\ref{fig:shifts4}b shows the repulsion of the eigenstates at $\bk_0 = (0.4, -0.4, 0.5)$, in reciprocal coordinates.  The crosses mark the values of $\bar{E}_{55}$ (red) and $\bar{E}_{58}$ (blue) expected in the absence of hybridization ($\approx P_{nn} E_n$). The level repulsion due to hybridization is shown along the arrows. 
As seen in Fig.~\ref{fig:Pmat2}b, the predominant non-zero off-diagonal element at $\bk_0$ is $P_{55,58}$, which indicates that the hybridization primarily involves only $\ket{A_{55}}$ and $\ket{A_{58}}$ as depicted by the green circles and lines in Fig.~\ref{fig:Pmat2}. Therefore, the repulsion of $\bar{E}_{55}$---the magnitude of which is given by the size of the arrows in the inset---can be analytically estimated by the perturbation formula as
 \[
\Delta \bar{E}_{55} = -\Delta \bar{E}_{58} \approx \frac{ |P_{55,58}|^2 E_{55} E_{58}}{P_{55,55}E_{55}- P_{58,58} E_{58}}= 0.1928\, \mbox{eV}\,,
\]
which in agreement to the actual value of 0.1381 eV.
\bigskip

All the eigenenergies seen in both panels (a) and (b) of Fig.~\ref{fig:shifts4}, \textit{i.e.} the TB subspace, are insensitive to any chosen value of $\kappa$ since the exact shifting matrix $Q_\mathcal{N}$ is used. The null subspace (eigenvalues not seen) have been rigidly shifted by $\kappa =2$ eV outside the region of interest. 
 In panel (c) we re-compute the TB Hamiltonian from (b) but using the approximated shifting matrix ($Q_\mathcal{N} \approx I_M - AA^\dag$) and $\kappa = 1.0$ eV. Different shifting patterns are observed: 
(\textit{i}) The eigenvalues of the null subspace shift rigidly with the value of $\kappa$ and, thus, are pinned along the horizontal line at 1.0 eV.
(\textit{ii}) The ``unhybridized'' bands of the TB subspace (black dots) show no noticeable difference with respect to panel Fig.~\ref{fig:shifts4}b. This confirms that while they formally acquire a $\kappa$ dependence, introduced by the approximated $Q_\mathcal{N}$, the effect is negligible for bands with high projectability.
(\textit{iii}) The high-projectability band $n$=55 (red dots) should also be insensitive to $\kappa$; nonetheless, it acquires a more noticeable dependence indirectly via its hybridization to the $\kappa$-dependent low-projectability band $n$=58. Consequently, the most noticeable changes of this band with respect to (b) 
\footnote{\label{note3} The TB matrix constructed using the approximated $Q_\mathcal{N}$ and $\kappa = 0$ eV yields the same non-null eigenvalues than when using the exact $Q_\mathcal{N}$ with any value of $\kappa$.}
happen around H where the hybridization is stronger.
(\textit{iv}) The low-projectability band $n$=58 (blue dots) shows a noticeable dependence on $\kappa$, especially around the lowest-projectability $\bk$ points. For instance, the value of $\kappa = 1.0$ eV effectively shifts the states around H by $\sim$ 0.9 eV, \textit{i.e.} from $\sim$ 0.4 eV [as in Fig.~\ref{fig:shifts4}(b)\footnotemark[5]] to $\sim$ 1.3 eV.

\section{Conclusions}
In this paper we have outlined a noniterative scheme to derive highly accurate \textit{ab-initio} TB Hamiltonian matrices in a minimal basis set representation.

Minimal basis sets may be insufficient to converge self-consistent quantum-mechanical calculations with linear combination of atomic orbitals, however, they are adequate for the purpose of projecting wavefunctions obtained with fully converged basis and building the reduced TB matrices.

Low-projectability Bloch states have spurious effects when included in the construction of the TB matrix. We have unambiguously shown the underlying mechanism for the formation of the spurious states. The removal of those states, via the application of a shifting matrix, delivers accurate TB matrices.

We have introduced an expression for the shifting matrix of a nonorthogonal set of vectors. This expression improves the quality of the TB Hamiltonian by removing any unwanted dependence that the shifting procedure had on the TB eigenstates of interest.

\appendix

\section{Projection of Bloch states on pseudo atomic orbitals}\label{sec:proj_bloch}
The pseudo-wavefunction Bloch state $\ket{\psit_{n\bk}}$ is expanded in plane-wave basis $\ket{\bk+\bG}$ as
\begin{equation}
\ket{\psit_{n\bk}}=\sum_{\bG} C_{\bG n \bk} \ket{\bk+\bG}.
\label{}
\end{equation}

The plane-wave basis
\begin{equation}
\langle \mathbf{r} | \mathbf{k+G} \rangle = \frac{1}{\sqrt{\Omega_0}} e^{i(\mathbf{k+G})\cdot \mathbf{r}}
\label{eq:PW}
\end{equation}
 is defined to be normalized to 1 over the volume of the primitive unit cell $\Omega_0$. The orthonormality of the basis $\braket{\bk+\bG}{\bk+\f{G'} }=\delta_{\bG \f{G'}} $ allows the expansion coefficients to be defined by the projection
\begin{equation}
C_{\bG n \bk} = \braket{\f{k+G}}{\psit_{n\bk}}
\label{eq:ccoefs}
\end{equation}

In the US/PAW pseudopotential formalisms, the projection of the all-electron (AE) wavefunctions $\psi_{n\mathbf{k}}$ onto an atomic orbital $\phi^{\bk}_{\mu}$ is computed in terms of their corresponding pseudized quantities $\psit_{n\mathbf{k}}$ and $\phit^{\bk}_{\mu}$, and the overlap operator $\hat{S}=\hat{1}+\sum_{I ij}| \beta^{\mathbf{k}}_{Ii}\rangle Q^I_{ij} \langle \beta^{\mathbf{k}}_{Ij}|$. The pseudo atomic orbitals (PAO) and beta projectors are defined as
\begin{equation}
\tilde{\phi}_{\mu}(\mathbf{r})  = R^{\phi}_{\mu}(r)Y^m_l(\widehat{\br})
\label{eq:phir}
\end{equation}
where $\mu \equiv \{Ilm\}$ is a composite index of the ion center $I$ and quantum numbers $\{lm\}$ of the PAO. The real-space beta projectors are analogously defined\cite{Ferretti_wannier_JPCM2007}, with $i \equiv\{lm\}$, as: 
\begin{equation}
\beta_{Ii}(\mathbf{r}) =
R^{\beta}_{Ii}(r)Y^m_l(\widehat{\br})
\label{}
\end{equation}

The localized basis $\ket{\phit^{\bk}_{\mu}}$ for periodic calculations is constructed from Bloch sums of the PAOs.

\begin{equation}
\braket{\br}{\phit^{\bk}_{\mu}}= \frac{1}{N} \sum_{\bR} e^{ i \bk \cdot \bR } \phit_{\mu}(\br -\f{\tau}_{\mu} - \bR ) ,
\label{eq:phikr}
\end{equation}
where $N$ is the number of lattice vectors $\bR$. The Bloch sum for the AO basis $\ket{\phi^{\bk}_{\mu}}$ follows the same definition. Notice that the factor $\frac{1}{N}$ implies normalization of $\ket{\tilde{\phi}^{\bk}_{\mu}}$ to 1 over the primitive unit cell, which is consistent with the normalization of the plane-wave basis in Eq.~\ref{eq:PW}.

Then, the projection coefficients are calculated in term of the pseudized quantities:
\begin{align}
B^\bk_{\mu n} &= \braket{\phi^{\mathbf{k}}_{\mu}}{\psi_{n\mathbf{k}}} = 
\langle\tilde{\phi}^{\mathbf{k}}_{\mu} |\hat{S}|\tilde{\psi}_{n\mathbf{k}}\rangle \notag\\ 
&= \langle\tilde{\phi}^{\mathbf{k}}_{\mu} | \tilde{\psi}_{n\mathbf{k}}\rangle +  
\sum_{\substack{\mathbf{G}\mathbf{G'}\\{Iij}}}
\langle\tilde{\phi}^{\mathbf{k}}_{\mu} | \beta^{\mathbf{k}}_{I i}\rangle Q^I_{ij} 
\langle\beta^{\mathbf{k}}_{I j}|\tilde{\psi}_{n\mathbf{k}}\rangle \, .
\end{align}

The integrals are more efficiently computed in the $\ket{\bG}$ basis. Using the identity $\hat{1}=\sum_\bG \ket{\bG} \bra{\bG} = \sum_{\mathbf{G}} \ket{\mathbf{k+G}}\bra{\mathbf{k+G}}$ one has

\begin{align*} 
B^\bk_{\mu n} &= \sum_{\mathbf{G}\mathbf{G'}} \langle\tilde{\phi}^{\mathbf{k}}_{\mu}| \mathbf{k+G}\rangle\langle\mathbf{k+G}|\tilde{\psi}_{n\mathbf{k}}\rangle + \\ 
&\sum_{\substack{\mathbf{G}\mathbf{G'}\\{I ij}}} 
\langle\tilde{\phi}^{\mathbf{k}}_{\mu} | \mathbf{k+G}\rangle\langle\mathbf{k+G}|\beta^{\mathbf{k}}_{I i}\rangle Q^I_{ij}
\langle\beta^{\mathbf{k}}_{I j}|\mathbf{k+G'}\rangle\langle\mathbf{k+G'}| \tilde{\psi}_{n\mathbf{k}}\rangle\, . 
\end{align*}

Using the definition in Eq.~\ref{eq:ccoefs} for the expansion coefficients:

\begin{align*} 
B^\bk_{\mu n} &= \langle \phi^{\mathbf{k}}_{\mu}| \psi_{n\mathbf{k}}\rangle = 
\sum_{\mathbf{G}} \langle\tilde{\phi}^{\mathbf{k}}_{\mu}| \mathbf{k+G}\rangle C_{\mathbf{G}n\mathbf{k}} + \notag \\
&\sum_{\substack{\mathbf{G}\mathbf{G'}\\{I ij}}} 
\langle\tilde{\phi}^{\mathbf{k}}_{\mu} | \mathbf{k+G}\rangle\langle\mathbf{k+G}|\beta^{\mathbf{k}}_{I i}\rangle Q^I_{ij}
\langle \beta^{\mathbf{k}}_{I j}|\mathbf{k+G'}\rangle C_{\mathbf{G'}n\mathbf{k}}\, , \label{eq:proj1}
\end{align*}
where the objects in brackets are given in Eqs.~\ref{eq:proj2} and \ref{eq:proj3}.

The coefficients $B^{\bk}_{\mu n}$ expand the Bloch state on a PAO basis $\{ \phi_\mu \} $. Furthermore, the coefficients on a L\"owdin orthonormal basis $\{ \bar{\phi}_\mu \}$ are readily obtained by    

\begin{equation}
\bar{B}^{\bk}_{\mu n} = \sum_{i}(S^{\bk}{}^{-\frac{1}{2}})_{i\mu} B^{\bk}_{\mu n},
\label{eq:proj4}
\end{equation}

where the upper bar symbol is used to indicate orthonormality and $S^{\bk}_{\mu \nu}= \braket{\phi^{\bk}_\mu}{\phi^{\bk}_\mu}  $ is the matrix of overlaps between PAOs.

L\"{o}wdin coefficients and orbitals are assumed throughout the main text  where we drop the upper bar and $\bk$ superscript in the notation of $\bar{B}^{\bk}$. 

\section{Projection of the plane-waves basis on pseudo atomic orbitals}
Using the relation $\hat{1}=\int \ket{\br}\bra{\br} d\br$ and Eqs.~\ref{eq:PW} and \ref{eq:phikr} to evaluate the projection $\langle \tilde{\phi}^{\mathbf{k}}_{\mu}| \mathbf{k+G}\rangle$ one has

\begin{align}
\braket{\phit^{\bk}_{\mu}}{ \bk+\bG} &=
 \int d\br \braket{ \phit^\bk_{\mu}}{ \br } \braket{\br}{\bk+\bG} \\
&= \frac{1}{N\sqrt{\Omega_0}} \sum_{\bR} e^{-i \bk \cdot \bR} \int d\br e^{i(\bk+\bG)\cdot \br} \phit_\mu^*(\br-\f{\tau}_{\mu} - \bR)\\
&= \frac{e^{i(\bk+\bG) \cdot \f{\tau}_\mu}}{N\sqrt{\Omega_0}} \sum_{\bR} e^{i \bG \cdot \bR} \int d\br e^{i (\bk+\bG)\cdot \br} \phit_\mu^*(\br)\\
&= \frac{e^{i(\bk+\bG) \cdot \f{\tau}_\mu}}{\sqrt{\Omega_0}} \int d\br e^{i (\bk+\bG)\cdot \br} \phit_\mu^*(\br) \, .
\end{align}

Using the plane-wave expansion $e^{i(\mathbf{k+G}) \cdot \br}= \sum_{l'm'} 4\pi i^{l'} j_{l'}(|\mathbf{k+G}| r) Y^{m'*}_{l'} (\widehat{\mathbf{k+G}}) Y^{m'}_{l'}(\widehat{\br})$, where the hat notation indicates the directional angles of the vector under it; Eq.~\ref{eq:phir}; and $d\br=r^2 \sin\theta dr d\theta d\varphi$, the last expression reduces to

\begin{align}
& \braket{\phit^{\bk}_{\mu}}{ \bk+\bG} = \frac{4\pi e^{i (\bk+\bG)\cdot \mathbf{\tau}_{\mu}}}{\sqrt{\Omega_0}} \sum_{l'm'} i^{l'} Y^{m'*}_{l'} (\widehat{\mathbf{k+G}}) \notag \\
& \times \int r^2 R^{\phi}_{\mu}(r) j_{l'}(|\mathbf{k+G}|r)dr \int  Y^{m*}_l (\widehat{\br}) Y^{m'}_{l'}(\widehat{\br}) \sin\theta d\theta d\varphi \, . \notag \\
\end{align}

With the the normalization identity $\int  Y^{m*}_l (\theta,\phi) Y^{m'}_{l'}(\theta,\phi) \sin\theta d\theta d\varphi = \delta_{ll'}\delta_{mm'}$, we arrive to the final expression:

\begin{equation}
\langle \tilde{\phi}^{\mathbf{k}}_{\mu}| \mathbf{k+G}\rangle = f_{\bG \mu \bk} Y^{m*}_{l} (\widehat{\mathbf{k+G}}) \int r^2 R^{\phi}_{\mu}(r) j_{l}(|\mathbf{k+G}|r)dr \, , 
\label{eq:proj2}
\end{equation}

with $f_{\bG \mu \bk}= 4\pi i^{l} {\Omega_0}^{-\frac{1}{2}} e^{i \mathbf{(k+G)}\cdot \mathbf{\tau}_{\mu}}$.%

Analogously for the projection on the beta functions:
\begin{equation}
\langle {\beta}^{\mathbf{k}}_{Ii}| \mathbf{k+G}\rangle = f_{\bG \mu \bk} Y^{m*}_{l} (\widehat{\mathbf{k+G}}) \int r^2 R^{\beta}_{Ii}(r) j_{l}(|\mathbf{k+G}|r)dr \, .
\label{eq:proj3}
\end{equation}

\section{Off-diagonal elements of the projector matrix}\label{sec:offdiag}
Given the projector matrix
\[
P_{nm} = (B^\dag B)_{nm} = \bra{\psi_n}\hat P\ket{\psi_m}\, ,
\]
since $\psi_n$ forms a complete Hilbert space, the matrix $P$ will also be a projection operator by closure.  Namely,
\[
P_{nm} = (P^2)_{nm} = \sum_j P_{nj}P_{jm} \, ,
\]
so there is a constraint for the important diagonal elements (and using the Hermitian nature of the matrix $P$) :
\[
p_n = p_n^2 + \sum_{m\ne n}|P_{nm}|^2 \, .
\]

This expression puts an upper bound of 1 on the diagonal elements. One can also define an upper bound on any off diagonal element via
\[
|P_{nm}| \le \min(\sqrt{p_n-p_n^2},\sqrt{p_m-p_m^2})\,.
\]

Notice that if $p_n=1$ (or $\approx 1$) which means perfect projection, then $P_{nm}=P_{mn}=0$ (or $\approx 0$) $\forall m\ne n$ so the entire $n$-th column and row of $P$ is zero (excluding the diagonal which is 1).

For cases of smaller projectability $p_n < 1$, each off diagonal entry will still be much smaller than $p_n$ since there are many of them in the sum rule; but the sum of their squares must add up to $p_n-p_n^2$.  
 
\acknowledgments
We want to thank Dr. Dmitri Volja for helpful discussions, the Texas Advanced Computing Center (TACC) at the University of Texas Austin for providing computing facilities, and the funding provided by the ONR-MURI under Contract No. N00014-13-1-0635. 
The authors acknowledge the Duke University Center for Materials Genomics for computational assistance.


\begin{thebibliography}{10}

\bibitem{Huckel1931Model}
E.~H{\"u}ckel, \emph{Quantentheoretische Beitr{\"a}ge zum Benzolproblem},
  Zeitschriftf{\"u}r Physik \textbf{70}, 628 (1931).

\bibitem{Jones1934Tight_binding}
H.~Jones, N.~F. Mott, and H.~W.~B. Skinner, \emph{A Theory of the Form of the
  X-Ray Emission Bands of Metals}, Phys.\ Rev. \textbf{45}, 379--384 (1934).

\bibitem{Papaconstantopoulos_JPCM_2003}
D.~A. Papaconstantopoulos and M.~J. Mehl, \emph{The Slater-Koster tight-binding
  method: a computationally efficient and accurate approach}, J.\ Phys.:\
  Conden.\ Matt. \textbf{15}, R413 (2003).

\bibitem{Goringe_TightBinding_RPP1997}
C.~M. Goringe, D.~R. Bowler, and E.~Hernandez, \emph{Tight-binding modelling of
  materials}, Reports on Progress in Physics \textbf{60}, 1447 (1997).

\bibitem{Colombo_Clusters_NATO1996}
L.~Colombo, \emph{Large Scale Simulations Using Tight Binding Molecular
  Dynamics}, in \emph{Large Clusters of Atoms and Molecules}, edited by
  T.~Martin (Springer Netherlands, 1996), \emph{NATO ASI Series}, vol. 313, pp.
  495--510, \doi{10.1007/978-94-009-0211-4_21}.

\bibitem{Heine_ES_LAE_SS1980}
V.~Heine, \emph{Electronic structure from the point of view of the local atomic
  environment}, Solid State Physics \textbf{35}, 1 (1980).

\bibitem{Andersen_muffintin_PRB2000}
O.~K. Andersen and T.~{Saha-Dasgupta}, \emph{Muffin-tin orbitals of arbitrary
  order}, Phys.\ Rev.\ B \textbf{62}, R16219--R16222 (2000).

\bibitem{Marzari2012MLWF}
N.~Marzari, A.~A. Mostofi, J.~R. Yates, I.~Souza, and D.~Vanderbilt,
  \emph{Maximally localized Wannier functions: Theory and applications}, Rev.\
  Mod.\ Phys. \textbf{84}, 1419--1475 (2012).

\bibitem{Lu2004QUAMBO}
W.~C. Lu, C.~Z. Wang, T.~L. Chan, K.~Ruedenberg, and K.~M. Ho,
  \emph{Representation of electronic structures in crystals in terms of highly
  localized quasiatomic minimal basis orbitals}, Phys.\ Rev.\ B \textbf{70},
  041101 (2004).

\bibitem{Agapito2015ACBN0}
L.~A. Agapito, S.~Curtarolo, and M.~{Buongiorno~Nardelli}, \emph{Reformulation
  of $\mathrm{DFT}+U$ as a Pseudohybrid Hubbard Density Functional for
  Accelerated Materials Discovery}, Phys.\ Rev.\ X \textbf{5}, 011006 (2015).

\bibitem{Horsfield1997Efficient_ab_initio_TB}
A.~P. Horsfield, \emph{Efficient \textit{ab-initio} tight binding}, Phys.\
  Rev.\ B \textbf{56}, 6594--6602 (1997).

\bibitem{Harris1985Simplified_method_for_the_energy}
J.~Harris, \emph{Simplified method for calculating the energy of weakly
  interacting fragments}, Phys.\ Rev.\ B \textbf{31}, 1770--1779 (1985).

\bibitem{Foulkes1989Tight_binding_models}
W.~M. Foulkes and R.~Haydock, \emph{Tight-binding models and density-functional
  theory}, Phys.\ Rev.\ B \textbf{39}, 12520--12536 (1989).

\bibitem{Sankey1989Multicenter_tight_binding_model}
O.~F. Sankey and D.~J. Niklewski, \emph{\textit{Ab initio} multicenter
  tight-binding model for molecular-dynamics simulations and other applications
  in covalent systems}, Phys.\ Rev.\ B \textbf{40}, 3979--3995 (1989).

\bibitem{Soler2002SIESTA}
J.~M. Soler, E.~Artacho, J.~D. Gale, A.~Garc{\'\i}a, J.~Junquera,
  P.~Ordej{\'o}n, and D.~S{\'a}nchez-Portal, \emph{The SIESTA method for ab
  initio order-N materials simulation}, J.\ Phys.:\ Conden.\ Matt. \textbf{14},
  2745 (2002).

\bibitem{Teichler1971Best_localized_Wannier_functions}
H.~Teichler, \emph{Best Localized Symmetry-Adapted Wannier Functions of the
  Diamond Structure}, Phys.\ Stat.\ Solidi\ B \textbf{43}, 307--318 (1971).

\bibitem{Note1}
\label {note2} Projection onto AO-like functions are used as the first
  iteration in the construction of MLWF functions. TB band structures obtained
  with these trial Wannier functions are equivalent to those from the
  construction in Eq.~\ref {eq:tb_hamiltonian}, see Ref.~\cite
  {Teichler1971Best_localized_Wannier_functions}.

\bibitem{Agapito_2013_projectionsPRB}
L.~A. Agapito, A.~Ferretti, A.~Calzolari, S.~Curtarolo, and
  M.~{Buongiorno~Nardelli}, \emph{Effective and accurate representation of
  extended Bloch states on finite Hilbert spaces}, Phys.\ Rev.\ B \textbf{88},
  165127 (2013).

\bibitem{Meyer2000LinearAlgebra}
C.~D. Meyer, \emph{Matrix analysis and applied linear algebra} (Siam, 2000).

\bibitem{Note2}
L. A. Agapito, A. Ferretti, S. Curtarolo, M. Buongiorno Nardelli, \protect
  \textit {in preparation}.

\bibitem{Note3}
\label {note1} This is possible as one can always compute more PW states to
  increase the number of columns of $A$. One can start with $M$ DFT states with
  the highest projectability $p_n$ to form $A$. If for some reason the
  corresponding $M$ columns vectors are not linearly independent, $A$ will not
  be invertible. In that case one can proceed in a organized iterative fashion:
  start with large pool of column vectors sorted by descending $p_n$,
  initialize a collection with the first vector and iteratively add another
  vector only if is linearly independent to the collection otherwise proceed to
  the next vector.

\bibitem{vasp}
G.~Kresse and J.~Furthm\"uller, \emph{Efficient iterative schemes for {\it ab
  initio} total-energy calculations using a plane-wave basis set}, Phys.\ Rev.\
  B \textbf{54}, 11169--11186 (1996).

\bibitem{quantum_espresso_2009}
P.~Giannozzi, S.~Baroni, N.~Bonini, M.~Calandra, R.~Car, C.~Cavazzoni,
  D.~Ceresoli, G.~L. Chiarotti, M.~Cococcioni, I.~Dabo, A.~{Dal Corso}, S.~{de
  Gironcoli}, S.~Fabris, G.~Fratesi, R.~Gebauer, U.~Gerstmann, C.~Gougoussis,
  A.~Kokalj, M.~Lazzeri, L.~Martin-Samos, N.~Marzari, F.~Mauri, R.~Mazzarello,
  S.~Paolini, A.~Pasquarello, L.~Paulatto, C.~Sbraccia, S.~Scandolo,
  G.~Sclauzero, A.~P. Seitsonen, A.~Smogunov, P.~Umari, and R.~M. Wentzcovitch,
  \emph{QUANTUM ESPRESSO: a modular and open-source software project for
  quantum simulations of materials}, J.\ Phys.:\ Conden.\ Matt. \textbf{21},
  395502 (2009).

\bibitem{PAW}
P.~E. Bl\"ochl, \emph{Projector augmented-wave method}, Phys.\ Rev.\ B
  \textbf{50}, 17953--17979 (1994).

\bibitem{PBE}
J.~P. Perdew, K.~Burke, and M.~Ernzerhof, \emph{Generalized gradient
  approximation made simple}, Phys.\ Rev.\ Lett. \textbf{77}, 3865--3868
  (1996).

\bibitem{Note4}
\label {note} The pseudo atomic orbitals used in this work are from public data
  sets:\\ \protect \texttt
  {http://www.quantum-espresso.org/pseudopotentials/}\\ C.pbe-n-kjpaw\protect
  \_psl.0.1.UPF \\ H.pbe-kjpaw\protect \_psl.0.1.UPF \\ \protect \texttt
  {http://qe-forge.org/gf/project/pslibrary/}\\ Co.pbe-n-kjpaw\protect
  \_psl.1.0.0.UPF \\ Sb.pbe-n-kjpaw\protect \_psl.1.0.0.UPF.

\bibitem{Curtarolo:art58}
W.~Setyawan and S.~Curtarolo, \emph{High-Throughput Electronic Structure
  Calculations: Challenges and Tools}, Comp.\ Mat.\ Sci. \textbf{49}, 299--312
  (2010).

\bibitem{CoSb_natmat:2015}
Y.~Tang, Z.~M. Gibbs, L.~A. Agapito, G.~Li, H.-S. Kim,
  M.~{Buongiorno-Nardelli}, S.~Curtarolo, and G.~J. Snyder, \emph{{Convergence
  of multi-valley bands as the electronic origin of high thermoelectric
  performance in CoSb$_3$ skutterudites}}, Nature\ Mater.  (2015).

\bibitem{Note5}
\label {note3} The TB matrix constructed using the approximated $Q_\protect
  \mathcal {N}$ and $\kappa = 0$ eV yields the same non-null eigenvalues than
  when using the exact $Q_\protect \mathcal {N}$ with any value of $\kappa $.

\bibitem{Ferretti_wannier_JPCM2007}
A.~Ferretti, A.~Calzolari, B.~Bonferroni, and R.~{Di~Felice}, \emph{Maximally
  localized Wannier functions constructed from projector-augmented waves or
  ultrasoft pseudopotentials}, J.\ Phys.:\ Conden.\ Matt. \textbf{19}, 036215
  (2007).

\end{thebibliography}
\end{document}